\documentclass[aps,preprint]{revtex4}
\usepackage{epsfig}

\begin{document}

\title{Subcritical series expansions for multiple-creation
nonequilibrium models}

\author{Carlos E. Fiore and M\'{a}rio J. de Oliveira}

\affiliation{Instituto de F\'{\i}sica\\
Universidade de S\~{a}o Paulo\\
Caixa Postal 66318\\
05315-970 S\~{a}o Paulo, S\~{a}o Paulo, Brazil}
\date{\today}

\begin{abstract}
Perturbative subcritical series expansion for the steady properties
of a class of one-dimensional nonequilibrium models characterized by
multiple-reaction rules are presented here. We developed  long series  
expansions for three nonequilibrium models: the pair-creation
contact process, the A-pair-creation contact process, which
is  closely related system to the previous model, and the triplet-creation
contact process. The long series allowed us to obtain accurate 
estimates for the critical point and critical exponents. 
Numerical simulations are also performed and compared with
the series expansions results.

PACS numbers: 02.50.Ga,05.10.-a,05.70.Ln
\end{abstract}
\maketitle

\section{Introduction}

Nonequilibrium systems have been used 
to describe a variety of problems in  physics, chemistry, biology 
and other areas. Different classes of nonequilibrium systems have 
been studied and a special 
attention has been devoted to systems with absorbing states.
The most studied system with absorbing states is the contact process (CP)
 \cite{harris}. The basic CP
is composed of spontaneous annihilation of particles and  creation of 
particles in empty sites provided they have at least one nearest neighbor site
occupied by a particle. The critical behavior of CP  belongs 
to the directed percolation (DP) universality class \cite{marro}.
Several approaches have been aplied  for describing 
the CP, such as numerical simulations \cite{marro}, 
continuous description by means of a Langevin equation  
\cite{dick1994,dornic},
renormalization group \cite{jef1,jef2} 
and series expansions \cite{dic89b,jen1993,jendic1994,innui}.
Very accurate estimates  for the critical
point and critical exponents have been obtained
for the basic CP model. However  for the models that concern us here, 
the  avaiable results  come  from only numerical simulations.
The use  of different tecniques may be a useful tool for  studying
the behavior of systems whose avaiable results are controversial.

In this paper, we developed  subcritical perturbative series expansions for
the steady state properties of 
a class of one-dimensional nonequilibrium models characterized by
catalytic creation of particles  in the presence of 
$m$-mers. We have considered here creation of particles in the presence of 
pairs of particles ($m=2$) and triplet of particles
($m=3$). To exemplify, we developed the series  expansion
for three nonequilibrium models:  the pair-creation
contact process (PCCP)
\cite{dictome,fiore1}, the A-pair-creation contact process (APCCP), which
is  closely related  to the PCCP, except that an empty site 
in the presence of at least a pair of particles
becomes occupied with rate $\lambda$ (in the PCCP it becomes occupied with
rate $\lambda/2$ times the number of pairs of adjacent particles),
and the triplet-creation
contact process (TCCP)\cite{dictome,fiore1}. 
Very precise estimates of the critical behavior and critical exponents
are obtained from the analysis of the series. We  compared our results 
with their  respective numerical simulations.  

\section{Operator formalism}

Let us consider a one-dimensional lattice with $N$ sites. 
The system  envolves in the time according to a Markovian
process with local and irreversible rules.   
The time evolution of the probability $P(\eta,t)$ of a given configuration
$\eta \equiv (\eta_{1},\eta_{2},...,\eta_{N})$ is given by the master
equation,
\begin{equation}
\label{master}
\frac{d}{dt} P(\eta,t)=\sum_{i}^{N}\{w_{i}(\eta^{i})P(\eta^{i},t)
-w_{i}(\eta)P(\eta,t)\},
\end{equation}
where   $\eta^{i} \equiv
(\eta_{1},\eta_{2},...,1-\eta_{i},...,\eta_{N})$.
The total rate $w_{i}(\eta)$  of the interacting models studied here 
is composed of two parts
\begin{equation}
w_{i}(\eta)= w_{i}^{a}(\eta)+ w_{i}^{c}(\eta),
\end{equation}
where the rates
$w_{i}^{a}(\eta)$ and $w_{i}^{c}(\eta)$ 
take into account the annihilation and the catalytic
creation of particles, respectively. 
The annihilation of a 
single particle is illustrated by the scheme $\bullet
\rightarrow \circ$, and the catalytic creation of
particles by $\circ \bullet \bullet
\rightarrow \bullet \bullet\bullet $ and  $\circ \bullet \bullet \bullet
\rightarrow \bullet \bullet \bullet \bullet $, 
for the PCCP and TCCP, respectively. 
The scheme for the APCCP is similar to the PCCP, but here if an empty
site has at least
one pair of adjacent particles, a new particle will be created with rate
$\lambda$. More precisely these rules are given by
\begin{equation}
w_{i}^{a}(\eta)=\eta_{i},
\end{equation}
for the annihilation subprocess, 
and $w_{i}^{c}(\eta)$ by
\begin{equation}
w_{i}^{c}(\eta)=\frac{\lambda}{2}(1-\eta_{i})(\eta_{i+1}\eta_{i+2}+\eta_{i-2}
\eta_{i-1}),
\end{equation}
for the PCCP,
\begin{equation}
w_{i}^{c}(\eta)=\lambda(1-\eta_{i})(\eta_{i+1}\eta_{i+2}+\eta_{i-2}
\eta_{i-1}-\eta_{i+1}\eta_{i+2}\eta_{i-2}
\eta_{i-1}),
\end{equation}
for the APCCP, and
\begin{equation}
w_{i}^{c}(\eta)=\frac{\lambda}{2}(1-\eta_{i})(\eta_{i+1}\eta_{i+2}\eta_{i+3}+
\eta_{i-3}\eta_{i-2}\eta_{i-1}),
\end{equation}
for the TCCP. 

Before developing the series expansion, it is necessary to write down
the master equation in terms of creation and annihilation operators. 
The base states corresponding to
a given site $i$ of the lattice are  $|\eta_{i}\rangle$
with $|\eta_{i}\rangle=|\circ \rangle$ or
$|\eta_{i}\rangle=|\bullet\rangle$
according to whether 
site  $i$ is vacant or occupied by a particle, respectively.
The creation  and annihilation  operators for the site $i$
are defined in the following manner
\begin{equation}
\label{creation}
A_{i}^{+}|\eta_{i}\rangle=(1-\eta_{i})|1-\eta_{i}\rangle,
\end{equation}
and
\begin{equation}
\label{annihilation}
A_{i}|\eta_{i}\rangle=\eta_{i}|1-\eta_{i}\rangle,
\end{equation}
and they satisfy the property $A_{i}^{+}A_{i}+A_{i}A_{i}^{+}=1$.

Introducing  the probability
vector $|\Psi(t)\rangle$ defined by
\begin{equation}
\label{psi}
|\Psi(t)\rangle=\sum_{\eta}P(\eta,t)|\eta\rangle,
\end{equation}
where $|\eta\rangle = \prod_{i=1}^{N} \otimes |\eta_{i} \rangle=
|\eta_{1},\eta_{2},...,\eta_{N}\rangle$ is  the vector defined
by the direct product of the base vectors.
Substituting  Eq. (\ref{psi}) in to  Eq. (\ref{master}) 
and using Eqs. (\ref{creation}) and (\ref{annihilation}),
the time evolution for the probability vector is given by,
\begin{equation}
\frac{d}{dt}|\Psi(t)\rangle =W|\Psi(t)\rangle,
\end{equation}
where the operator $W=W_0 + \lambda V$ is a sum of 
the unperturbed term $W_0$ and a perturbed term $\lambda V$. 
The operator
$W_{0}$, that takes into account only the annihilation subprocess, 
is a nontinteraction term, given by
\begin{equation}
\label{wo}
W_{0}=\sum_{i}(A_{i}-A_{i}^{+}A_{i}).
\end{equation}
Each term of the summation
has the following set of  right and left eigenvectors
\begin{equation}
|0\rangle \equiv |\circ \rangle , \hspace{1 cm}\langle  0| 
\equiv \langle \circ| + \langle \bullet|,
\end{equation}
with eigenvalue $\Lambda_{0}=0$ and
\begin{equation}
|1\rangle \equiv -|\circ \rangle + |\bullet \rangle, 
\hspace{1 cm}\langle  1| \equiv \langle \bullet|, 
\end{equation}
with eigenvalue $\Lambda_{1}=-1$.
The operator $V$, correspondint to the catalytic creation of
particles, is an interacting term, given by
\begin{equation}
V=\frac{1}{2}
\sum_{i}S_{i}(n_{i+1}n_{i+2}+n_{i-2}n_{i-1}),
\end{equation}
for the PCCP,
\begin{equation}
\label{V}
V= \sum_{i}S_{i}(n_{i+1}n_{i+2}+n_{i-2}n_{i-1}
-n_{i+1}n_{i+2}n_{i-2}n_{i-1}),
\end{equation}
for the APCCP, and 
\begin{equation}
V=\frac{1}{2}
\sum_{i}S_{i}(n_{i+1}n_{i+2}n_{i+3}+n_{i-3}n_{i-2}n_{i-1}),
\end{equation}
for the TCCP, where the operator $S_{i}$ is given by 
$S_{i}=A_{i}^{+}-A_{i}A_{i}^{+}$ and $n_{i}=A_{i}^{+}A_{i}$ is the 
operator number.

To  find the steady vector $|\psi\rangle$, that satisfies
the steady condition
$(W_0+\lambda V)|\psi\rangle=0$, we assume that
\begin{equation}
\label{perturbative}
|\psi\rangle=|\psi_{0}\rangle 
+\sum_{\ell=1}^{\infty}\lambda^{\ell}|\psi_{\ell}\rangle,
\end{equation}
where $|\psi_{0}\rangle $  is the steady solution
of the non-interacting term $W_{0}$ satisfying the stationary condition
\begin{equation}
W_{0}|\psi_{0}\rangle=0.
\end{equation}
The vectors $\psi_{\ell}\rangle$ can be generated recursively
from the  initial state $|\psi_{0}\rangle$. 
Following Dickman \cite{dic89b}, we get the following
recursion relation
\begin{equation}
\label{recursive}
|\psi_{\ell}\rangle=-RV|\psi_{\ell-1}\rangle.
\end{equation}
The operator $R$ is the inverse of $W_{0}$ in the subspace of
vectors with nonzero eigenvalues and given by
\begin{equation}
\label{R}
R=\sum_{n(\neq 0)}|\phi_{n}\rangle \frac{1}{\Lambda_{n}}\langle  \phi_{n}|,
\end{equation}
where $|\phi_{n}\rangle$ and $\langle  \phi_{n}|$  are 
right and left eigenvectors of $W_{0}$,
respectively, with nonzero eigenvalue $\Lambda_{n}$. 

We notice that the steady solution of the noninteracting operator $W_{0}$
corresponds to the vacuum   
$|\psi_{0}\rangle=|\phi_{0}\rangle=|.0.\rangle$.
Since the creation of particles is catalytic, then if we start from the vacuum 
state, we will obtain a trivial steady vector namely 
$|\psi \rangle=|\psi_{0} \rangle$.
To overcome this problem, it is necessary to introduce a modification
on the rules of the models. The necessity of introducing a small 
modification on systems with absorging states in order
to get nontrivial steady states 
has been considered previously by Tom\'e and de Oliveira
\cite{tome2005} and  by de Oliveira \cite{oliveira1}.

\section{Generating the subcritical series}

The  modification we have made consists in introducing 
a spontaneous creation of particles in two specified adjacent sites
for the PCCP and APCCP. The chosen sites are $i=0$ and
$i=1$, so that the rates $w_{0}^{a}(\eta)$ and $w_{1}^{a}(\eta)$ 
are changed to
\begin{equation} 
w_{0}^{a}(\eta)=(1-q)\eta_{0}+q(1-\eta_{0}),
\end{equation}
and
\begin{equation} 
w_{1}^{a}(\eta)=(1-q)\eta_{1}+q(1-\eta_{1}),
\end{equation}
where $q$ is supposed to be a small parameter.
This modification leads to the following expression to the operator $W_{0}$
\begin{equation} 
W_{0}=\sum_{i} W_{0i}
+ q(S_{00}+S_{01}-W_{00}-W_{01}).
\end{equation}
The steady state $|\psi_{0}\rangle$ of $W_0$ is not
the vacuum state anymore. Now, it is given by 
\begin{equation}
|\psi_{0}\rangle=|.0.\rangle+
2q|.10.\rangle+q^{2}|.11.\rangle,
\end{equation}
where all sites before and after the symbol ``.'' are empty.

Two remarks are in order. First, only the last term in
$|\psi_{0}\rangle$ will give nonzero contributions to the expansion
so that $|\psi_\ell \rangle$, $\ell\geq 1$, will be of the
order $q^2$. Second, 
although the change in $W_{0}$ will cause a
change in $R$, only the terms of zero order in the expansion in $q$, 
given by the right-hand side of Eq. (\ref{R}), will be necessary
since the corrections in $R$ will contribute to terms of order larger than
$q^2$. For instance, the two first vectors, $|\psi_{1} \rangle$ and $|\psi_{2}
\rangle$, for the PCCP are given by
\begin{equation}
\label{vector1}
|\psi_{1}\rangle=q^{2}\{2|.1.\rangle+
|.11.\rangle+
|.101.\rangle+
\frac{2}{3}|.111.\rangle \},
\end{equation}
and
\begin{equation}
|\psi_{2}\rangle=q^{2}\{\frac{2}{3}|.1.\rangle+
\frac{1}{3}|.11.\rangle+
\frac{1}{3}|.101.\rangle
+\frac{2}{9}|.111.\rangle+
\frac{2}{3}|.1001.\rangle+
\frac{4}{9}|.1101.\rangle
+\frac{4}{9}|.1011.\rangle
+\frac{1}{3}|.1111.\rangle\},
\end{equation}
The translational invariance of the system is assumed.

For the TCCP, 
the rates $w_i^{a}(\eta)$, $i=0,1,2$
are modified similarly and
an analogous initial vector  $|\psi_{0}\rangle$ is obtained.
However, the vectors 
$|\psi_\ell \rangle$, $\ell\geq 1$, will be of the
order $q^3$.

The series expansions for the probability vector $|\psi \rangle$ obtained
here are equivalent to 
the Laplace transform $|\tilde \Psi(s)\rangle$ of the time dependent
vector probability $|\Psi(t)\rangle$ in the subcritical regime. If we
assume that  $|\tilde \Psi(s)\rangle$ can be expanded in powers 
of $\lambda$,
\begin{equation}
|\tilde \Psi(s)\rangle=|\tilde \Psi_{0}\rangle+\lambda |\tilde \Psi_{1}\rangle+
\lambda^{2} |\tilde \Psi_{2}\rangle+\ldots,
\end{equation} 
where 
\begin{equation}
|\tilde \Psi_{0}\rangle=(s-W_{0})^{-1}|X_0\rangle,
\end{equation} 
and
\begin{equation}
|\tilde \Psi_{\ell}\rangle=(s-W_{0})^{-1}V|\tilde \Psi_{\ell-1}\rangle,
\end{equation} 
where $|X_0\rangle=|\bullet\bullet\rangle$ for the PCCP and APCCP and 
$|X_0\rangle=|\bullet\bullet\bullet\rangle$ for the TCCP. 
The two first vectors for the PCCP is given by
\begin{equation}
\label{laplace0}
|\tilde \Psi_{0}\rangle=
\frac{1}{s}|.00.\rangle+
2s_{1}|.10.\rangle+s_{2}|.11.\rangle,
\end{equation}
and
\begin{equation}
\label{laplace1}
|\tilde \Psi_{1}\rangle=2s_{2}(
s_{1}|.1.\rangle+
s_{2}|.11.\rangle+
s_{2}|.101.\rangle+
s_{3}|.111.\rangle),
\end{equation}
where $s_{r}=1/(s+r)$.
In the limit $s \rightarrow 0$, Eq. (\ref{laplace1}) becomes identical
(by a factor $2q^2$) to  the Eq. (\ref{vector1}), that is
$|\tilde \Psi_{1}\rangle =|\psi_{1}\rangle/2q^2$.
The next orders of the expansion will 
also produce vectors that follows a similar relationship,
namely $|\tilde \Psi_{\ell}\rangle =|\psi_{\ell}\rangle/2q^2$.
Therefore, the steady-state vector $|\Psi\rangle$ has
a close relationship with
the Laplace transform $|\tilde \Psi(s)\rangle$ of the time dependent
vector probability $|\Psi(t)\rangle$ in the subcritical regime.

\begin{table}
\caption{Coefficients for the series expansion for total number of
particles $N$ corresponding to the PCCP, APCCP, and TCCP}
\smallskip
\begin{tabular}{|c|c|c|c|} 
\hline
\emph{n} & PCCP &APCCP&TCCP \\
\hline
 0 &   2.00000000000000 $\times 10^{0}$&  2.00000000000000 $\times 10^{0}$&
 3.00000000000000 $\times 10^{0}$\\
  1 &   2.00000000000000 $\times 10^{0}$&  2.00000000000000$\times 10^{0}$&
2.00000000000000$\times 10^{0}$\\
  2 &   6.66666666666667 $\times 10^{-1}$& 6.66666666666667 $\times 10^{-1}$&
6.66666666666667 $\times 10^{-1}$\\
  3 &   2.22222222222222 $\times 10^{-1}$& 2.22222222222222 $\times 10^{-1}$&
2.22222222222222 $\times 10^{-1}$\\
  4 &   1.48148148148148 $\times 10^{-1}$& 1.14814814814814 $\times 10^{-1}$&
2.07407407407407$\times 10^{-1}$\\
  5 &  -1.97530864197531 $\times 10^{-2}$&-5.86419753086417$\times 10^{-3}$&
3.80246913580247$\times 10^{-2}$\\
  6 &   4.46913580246914 $\times 10^{-2}$& 2.88117283950617$\times 10^{-2}$&
 -1.83938859494415$\times 10^{-2}$\\
  7 &  -1.57722908093278 $\times 10^{-2}$&-8.63692925729961$\times 10^{-3}$&
 4.64386215391508$\times 10^{-2}$\\
  8 &   9.53583512966229 $\times 10^{-3}$& 4.98937532660526$\times 10^{-3}$&
  2.17580715092230$\times 10^{-2}$\\
  9 &  -3.32566769780614 $\times 10^{-3}$&-6.44486162434792$\times 10^{-4}$&
 -9.02958001361142$\times 10^{-2}$\\
 10 &   2.47853920668470 $\times 10^{-3}$&-5.81533882116271$\times 10^{-4}$&
 1.45054908178555$\times 10^{-1}$\\
 11 &  -2.71937552830685 $\times 10^{-3}$& 1.24197428649364$\times 10^{-3}$&
  -1.65481868690147$\times 10^{-1}$\\
 12 &   3.41451431396303 $\times 10^{-3}$&-1.49842574202289$\times 10^{-3}$&
 1.63724920138393$\times 10^{-1}$\\
 13 &  -3.83526968827333 $\times 10^{-3}$& 1.67871776734916$\times 10^{-3}$&
-1.52183784122793$\times 10^{-1}$\\
 14 &   3.98069888064335 $\times 10^{-3}$&-1.79354081284144$\times 10^{-3}$&
 1.38404751317555$\times 10^{-1}$\\
 15 &  -3.93493438614195 $\times 10^{-3}$& 1.85786747449938$\times 10^{-3}$&
-1.21849140331913$\times 10^{-1}$\\
 16 &   3.80842329596270 $\times 10^{-3}$&-1.87771422345257$\times 10^{-3}$&
9.93112122560991$\times 10^{-2}$\\
 17 &  -3.66125398764534 $\times 10^{-3}$& 1.86718390972607$\times 10^{-3}$&
 -6.89396479237052$\times 10^{-2}$\\
 18 &   3.51794756163694 $\times 10^{-3}$&-1.83825949160899$\times 10^{-3}$&
 3.19324394323512$\times 10^{-2}$\\
 19 &  -3.38275717362883 $\times 10^{-3}$& 1.79908752660346$\times 10^{-3}$&
8.25261267789745$\times 10^{-3}$\\
 20 &   3.25366363586711 $\times 10^{-3}$&-1.75411933206063$\times 10^{-3}$&
-4.75544305218625$\times 10^{-2}$\\
 21 &  -3.12851358927982 $\times 10^{-3}$& 1.70572551425842$\times 10^{-3}$&
 8.27545456947378$\times 10^{-2}$\\
 22 &   3.00686185792610 $\times 10^{-3}$&-1.65537108002048$\times 10^{-3}$&
-1.12219860616667$\times 10^{-1}$\\
 23 &  -2.88968721395594 $\times 10^{-3}$& 1.60418342715199$\times 10^{-3}$&
1.36044569387758$\times 10^{-1}$\\
24 &   2.77858827033122 $\times 10^{-3}$&-1.55308848701112$\times 10^{-3}$&
-1.55881872792929$\times 10^{-1}$\\
 25 &  -2.67506636244736 $\times 10^{-3}$& 1.50279564792145$\times 10^{-3}$&
 1.74670507433362$\times 10^{-1}$\\
 26 &   2.58007455475132 $\times 10^{-3}$&-1.45378346183541$\times 10^{-3}$&
-1.96296951393837$\times 10^{-1}$\\
\hline
\end{tabular}
\label{coef}
\end{table}

\begin{table}
\caption{Continued from Table \ref{coef}}
\smallskip
\begin{tabular}{|c|c|c|c|} 
\hline
 27 &  -2.49382610216520 $\times 10^{-3}$& 1.40632844965132$\times 10^{-3}$&
 2.25163191741039$\times 10^{-1}$\\
 28 &   2.41581374222586 $\times 10^{-3}$&-1.36056168112988$\times 10^{-3}$&
 -2.65667885365566$\times 10^{-1}$\\
 29 &  -2.34497974277679 $\times 10^{-3}$& 1.31652716913818$\times 10^{-3}$&
3.21687696957426$\times 10^{-1}$\\
 30 &   2.27996894225733 $\times 10^{-3}$&-1.27422570713090$\times 10^{-3}$&
-3.96199475827942$\times 10^{-1}$\\
 31 &  -2.21939383143445 $\times 10^{-3}$& 1.23364009587133$\times 10^{-3}$&
 4.91189226879101$\times 10^{-1}$\\
 32 &   2.16205054135610 $\times 10^{-3}$&-1.19474540609862$\times 10^{-3}$&
-6.07957756683959$\times 10^{-1}$\\
 33 &  -2.10704796775453 $\times 10^{-3}$& 1.15751047699458$\times 10^{-3}$&
 7.47871824764928$\times 10^{-1}$\\
 34 &   2.05384177252338 $\times 10^{-3}$&-1.12189614019704$\times 10^{-3}$&
--------\\
 35 &  -2.00219063554985 $\times 10^{-3}$& 1.08785366190089$\times 10^{-3}$&
--------\\
 36 &   1.95206680869935 $\times 10^{-3}$&-1.05532488699381$\times 10^{-3}$&
--------\\
 37 &  -1.90355504280079 $\times 10^{-3}$& 1.02424413320142$\times 10^{-3}$&
--------\\
 38 &  1.85676625916399 $\times 10^{-3}$& -9.94541141539853$\times 10^{-4}$&
--------\\ 
\hline
\end{tabular}
\end{table}

\section{Analysis of the series}

To calculate the coefficients of $|\Psi_\ell\rangle$ 
in the base $|\eta\rangle$ we have built a computational algorithm
to take account of all configurations.
The configuration can be expressed in terms of a binary number 
$\eta_1 +\eta_2 2 +\eta_3 2^2+ \ldots$ 
representing the vector $|\eta\rangle$.
For example, the binary number 1101 corresponds to the configuration
 $|.1101.\rangle $ and we need to store
only the value of the coefficient of 1101. 
By this procedure we were able to determine
the coefficients of all vectors $|\Psi_\ell\rangle$
up to the 26th order in $\lambda$ 
for the PCCP (and APCCP) and to the 25th order for the TCCP.

From the series expansion of the vector $|\psi \rangle$, it is possible
to determine several quantities, such as survival probability, the total number
of particles, and the correlation function. 
In this paper, however, we will be concerned only with the series 
expansion for the total number of particles $N$, given by
\begin{equation}
N=\langle .0.|\sum_{i}n_{i}| \psi \rangle.
\end{equation}
One can show that the coefficient of
$\lambda^\ell$ in the expansion for $N$ is simply the coefficient of
$|.1.\rangle$ in $|\psi_\ell\rangle$. This allows us
to get a longer series for the number of particles. For the PCCP and APCCP
we obtained 38 terms and for the TCCP we obtained 33 terms.
The resulting series for the total number of particles
of the three models considered here are listed in Table \ref{coef}.

From the series expansion of a given quantity, in the present case, $N$, 
we can determine the critical point and its corresponding critical exponent
by means of a Pad\'e analysis.
Since the series developed here is related to the Laplace transform
of the total number of particles, both will have the same
critical behavior namely \cite{jen1993,jendic1994}
\begin{equation}
N \sim (\lambda -\lambda_{c})^{-\nu_{\,||}(1+\eta)},
\end{equation}
where $\nu_{\,||}$ and $\eta$ are the exponents related to the 
time correlation length and to the growth
of the number of particles, respectively.

A preliminar analysis is done by  performing unbiased estimates for 
determining both the critical point $\lambda_{c}$ and the critical exponent
by means of the Pad\'e aproximants \cite{guttmann, baker}.
This approach consists of analysing the
serie $(d/d \lambda)$ ${\rm ln} N$ by a Pad\'e approximant.
The critical exponent and the critical parameter $\lambda_c$
are obtained from the pole and the residue at this pole,
respectively.
We have obtained unbiased analysis for the three models considered here.
However, they  give us estimates that does not seem to
improve significatively when we consider higher-order Pad\'e approximants.
For example, for the PCCP  
the approximant [13/13]  gives $\lambda_{c}=7.62$ and
$\nu_{\,||}(1+\eta)=2.71$ whereas the approximant 
[16/16] gives $\lambda_{c}=7.54$ and $\nu_{\,||}(1+\eta)=2.56$. 

\begin{figure}
\centering
\epsfig{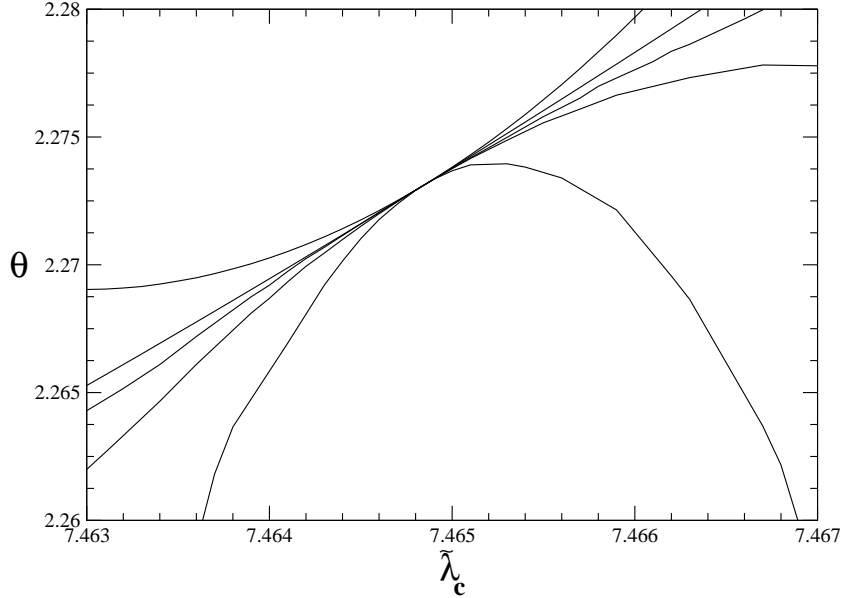}
\vspace{0.1 cm}
\caption{Biased estimates of $ \theta=\nu_{\,||}(1+\eta)$ as
a function of $\tilde\lambda_{c}$ derived from the Pad\'e approximants
to the series $(\lambda-\tilde\lambda_{c})(d/d \lambda)$ 
$\ln N=\theta$ evaluated at $\tilde\lambda_{c}$ for the PCCP.
The approximants shown are [17/18], [18/17], [18,19], [19/18], and [18/18].}
\label{padepccp}
\end{figure}

\begin{figure}
\centering
\epsfig{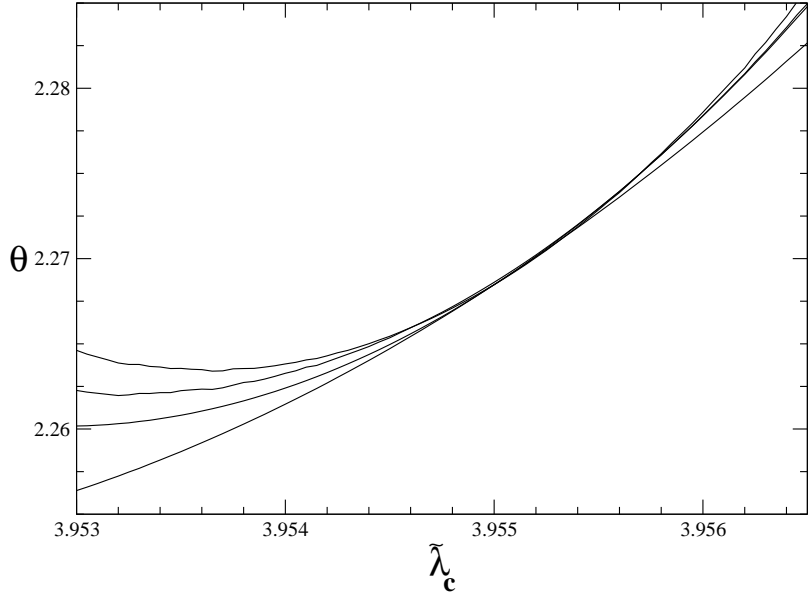}
\caption{Biased estimates of $ \theta=\nu_{\,||}(1+\eta)$ as
a function of $\tilde\lambda_{c}$ derived from the Pad\'e approximants
 to the series 
$(\lambda-\tilde\lambda_{c})(d/d\lambda)$ 
$\ln N=\theta$ evaluated at $\tilde\lambda_{c}$ for the APCCP.
The approximants shown are [18/18], [17,18], [18/19], and [19/18].}
\label{padeapccp}
\end{figure}

\begin{figure}
\setlength{\unitlength}{1.0cm}
\includegraphics[scale=0.45]{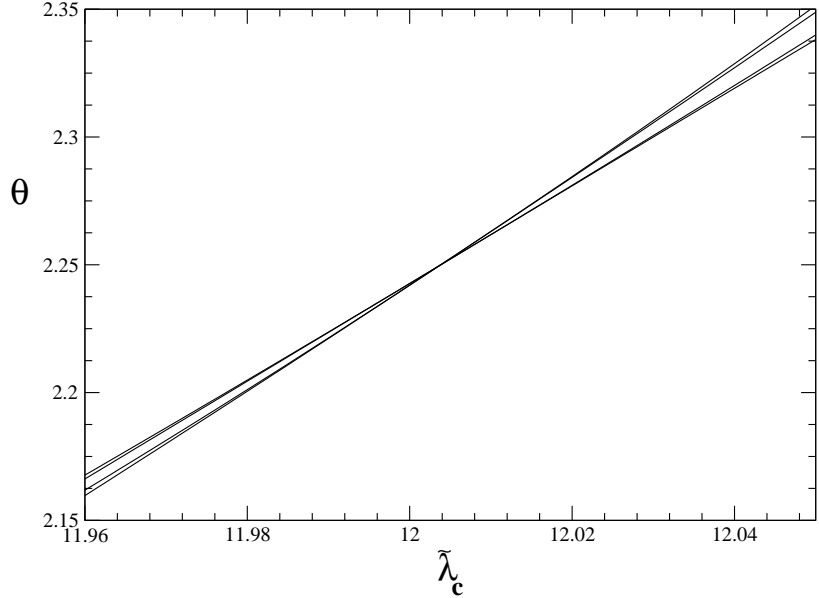}
\caption{Biased estimates of $ \theta=\nu_{\,||}(1+\eta)$ as
a function of $\tilde\lambda_{c}$ derived from the Pad\'e approximants
to the series 
$(\lambda-\tilde\lambda_{c})(d/d \lambda)$ 
$\ln N=\theta$ evaluated at $\tilde\lambda_{c}$ for the TCCP.
The approximants shown are [15/15], [15,16], [16,15], and [16,16].}
\label{padetccp}
\end{figure}

Much more reliable estimates are obtained when we perform 
biased analysis,
which is set up by looking at Pad\'e aproximants to the series
$(\lambda-\lambda_{c})(d/d \lambda)$ ${\rm ln} P=\theta$
\cite{jen93b,jen1991,well}. For
a trial value  of $\tilde \lambda_{c}$, we develop the serie above obtaining
$\theta(\tilde\lambda_{c})$ for a given Pad\'e aproximant $[m/n]$.
We can build curves for different Pad\'e approximants by repeating this
procedure for several trials $\tilde \lambda_{c}$
and we expect that they intercept at  the critical  
point ($ \lambda_{c}$, $ \theta(\lambda_{c})$).
In the Figs. \ref{padepccp}, \ref{padeapccp} and \ref{padetccp}
we plotted the curves obtained
by considering different Pad\'e aproximants for the three models.

From the Figs. \ref{padepccp}, \ref{padeapccp} and \ref{padetccp},
we see a very narrow intersection
of the Pad\'e approximants, revealing the utility of this approach.
However, as pointed out by Guttmann \cite{guttmann}, it is difficult to
estimate uncertainties in series calculations. Thus,
in order to give a more  realistic estimate of the 
quantities measured here and their associated uncertainties, 
we have estimated them by taking into account the first and last
crossings among various Pad\'e approximants. 
The values of the critical parameters obtained for the three models
are summarized in the Table \ref{table1}.
The estimates of $\lambda_c$ for the PCCP and TCCP are in excellent agreement
with the corresponding values $\lambda_{c}=7.464(2)$ and 
$\lambda_{c}=12.00(1)$ obtained from numerical simulations 
\cite{dictome,fiore1}. 

\begin{table}
\label{table1}
\caption{Biased estimates for $\lambda_{c}$ and $\nu_{\,||}(1+\eta)$
from the Pad\'e approximants for the three models considered here
together with the values for the basic contact process (CP)
\cite{jen93b}.}
\begin{tabular}{|c|c|c|} 
\hline
Model & $\lambda_{c}$ & $\nu_{\,||}(1+\eta)$\\
\hline 
PCCP  &  7.4650(6) & 2.274(3) \\
APCCP &  3.9553(5) & 2.272(4) \\
TCCP  & 12.01(2)   & 2.26(2)  \\
CP    &  3.29782   & 2.2772   \\
\hline
\end{tabular}
\end{table}

\begin{figure}
\setlength{\unitlength}{1.0cm}
\includegraphics[scale=0.5]{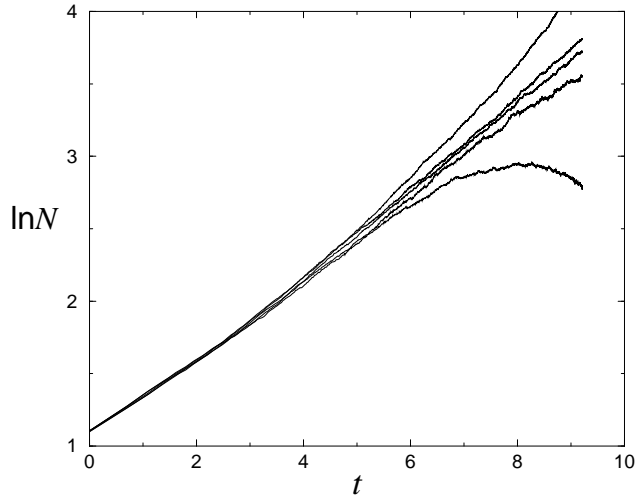}
\caption{Log-log plot of the number of particles $N_{s}$ versus the
time $t$ for some values of $\lambda$ for the APCCP. 
From top to bottom
$\lambda=3.97$, $3.9558$, $3.9553$, $3.95$, and $3.93$.}
\label{apccpnp}
\end{figure}

\section{Numerical simulation}

As a check of the accuracy of the results obtained here, we
have performed spreading simulations for the APCCP, since its
critical point is unknown in the literature.
Following Grassberger and de la Torre \cite{grassberger}, we
started from a initial configuration close to the absorbing state 
with only two adjacent particles, we can study the time evolution 
the survival probablility $P_{s}(t)$, the mean number of particles $N_{s}(t)$,
and the mean-square distance of the particles from the origin  $R(t)$.
At the critical point, these quantities are governed by power-laws whose their
related critical exponents are named $\delta$, $\eta$ and $z$,
respectively. 
Off-critical point, we expect deviations from the power-law behavior.
In the Fig.  \ref{apccpnp},
we plotted the quantity
$N_{s}(t)$  versus the time $t$ for some values of
$\lambda$. Analogous analysis can be done for determing the exponents
$\delta$ and $z$.
At the critical value $\lambda_{c}=3.9553$, our data for the
three quantities 
$N_{s}(t)$, $P_s(t)$, and $R(t)$ 
follow indeed a power law behavior, whose critical exponents
are consistent with those of the DP universality class.

\section{Conclusion}

We have derived subcritical series expansions for studying
the critical behavior of three  nonequilibrium systems characterized 
by multiple-creation of particles.
Although series expansion have been applied  sucessfully for the contact 
process and similar models \cite{jen93b,jen1993,jendic1994}, this is the first
time that these systems, with multi-reaction rules, has been treated 
by means of a technique other than numerical simulations. 
With exception of the TCCP, whose value of $\lambda_c$ are in the same
level of precision of numerical simulation estimates,
the subcritical series expansion give us the
best estimates for the critical point of the models considered here.
The critical exponents are consistent with those 
related to models beloging in the DP universality class.
We remark finally that the present approach may be very useful to determine
the critical behavior and universality classes for other
nonequilibrium systems. 

\section*{Acknowledgement}

We acknowledge W. G. Dantas for his critical reading of the
manuscript and one of us (C. E. Fiore) acknowledges the financial 
support from  Funda\c{c}\~ao de Amparo \`a Pesquisa do
Estado de S\~ao Paulo (FAPESP) under Grant No. 03/01073-0.



\begin{thebibliography}{99}

\bibitem{harris} T. E. Harris, Ann. Probab. {\bf 2}, 969 (1974).

\bibitem{marro}  J. Marro and R. Dickman, {\it Nonequilibrium Phase 
Transitions in Lattice Models} (Cambridge University Press, 
Cambridge, 1999).


\bibitem{dick1994}  R. Dickman,  Phys. Rev. E {\bf 50},
4404 (1994).

\bibitem{dornic} I. Dornic, H. Chat\'e, and M. A. Mu\~noz,
Phys. Rev. Lett. {\bf 94}, 100601 (2005).

\bibitem{jef1} J. Hooyberghs and C. Vanderzande, J. Phys. A {\bf 33},
907 (2000).

\bibitem{jef2}  J. Hooyberghs and C. Vanderzande,
Phys. Rev. E, {\bf 63}, 041109 (2001).

\bibitem{dic89b}  R. Dickman, J. Stat. Phys. {\bf 55}, 997 (1989).

\bibitem{jen1993}  I. Jensen and R. Dickman, 
J. Phys. A {\bf 26}, L151 (1993).

\bibitem{jendic1994} I. Jensen and R. Dickman, Physica A {\bf 203},
175 (1994).

\bibitem{innui}
N. Inui and A. Yu. Tretyakov, Phys. Rev. Lett {\bf 80}, 5148 (1998).

\bibitem{dictome} R. Dickman and T. Tom\'e, Phys. Rev. A {\bf 44},
4833 (1991).

\bibitem{fiore1} C. E. Fiore and M. J. de Oliveira, Phys. Rev. E, {\bf 70},
046131 (2004).

\bibitem{tome2005}  T. Tom\'e and M. J. de Oliveira, Phys. Rev. E {\bf 72},
026130 (2005).

\bibitem{oliveira1} M. J. de Oliveira (unpublished).

\bibitem{guttmann} {\it Phase Transitions and Critical Phenomena}, vol. 13,
edited by C. Domb and J. L. Lebowitz (Academic Press, New York, 1989).

\bibitem{baker} G. A. Baker, {\it Quantitative Theory of Critical Phenomena}, 
vol. 3, edited by C. Domb and J. L. Lebowitz (Academic Press, Boston, 1990). 

\bibitem{jen93b}   I. Jensen and R. Dickman, J. Stat. Phys.
{\bf 71}, 89 (1993).

\bibitem{jen1991}   R. Dickman and I. Jensen, 
Phys. Rev. Lett {\bf 67}, 2391 (1991).

\bibitem{well} W. G. Dantas and J. F. Stilck, J. Phys. A, 
{\bf 38}, 5841 (2005).

\bibitem{grassberger} P. Grassberger and A. de la Torre, Ann. Phys. (N.Y.)
{\bf 122}, 373 (1979).

\end{thebibliography}
\end{document}